\documentclass[%
 preprint,
 amsmath,amssymb,
 citeautoscript,
]{revtex4-2}

\usepackage[pdftex]{graphicx} % dvipdfmx driver for Japanese
\usepackage{dcolumn}
\usepackage{bm}
\usepackage{amsmath}
\usepackage{mathtools}%\coloneqq
\usepackage[version=4]{mhchem}
\usepackage{subcaption}%subfigure
\usepackage{physics}
\usepackage[utf8]{inputenc} % or other encoding
\usepackage[T1]{fontenc}
\usepackage{hyperref} % ensure hyperref is compatible with dvipdfmx
\usepackage{CJK} %font
\usepackage{multirow}
\usepackage{float} 
\hypersetup{
    colorlinks=true,
    linkcolor=blue,
    citecolor=blue,
    filecolor=blue, 
    urlcolor=blue
}
\setcitestyle{sort&compress,open={},close={},numbers,super}

\makeatletter
\renewcommand{\section}{\@startsection{section}{1}{\z@}%
  {-3.5ex \@plus -1ex \@minus -.2ex}%
  {2.3ex \@plus.2ex}%
  {\normalfont\bfseries}}
\renewcommand{\subsection}{\@startsection{subsection}{2}{\z@}%
  {-3.25ex \@plus -1ex \@minus -.2ex}%
  {1.5ex \@plus .2ex}%
  {\normalfont\slshape}}
\makeatother

\usepackage{setspace}
\usepackage{titlesec}
\titlespacing*{\section}
{0pt}{2\baselineskip}{1\baselineskip}

\titlespacing*{\subsection}
{10pt}{1.3\baselineskip}{0.5\baselineskip}
\begin{document}

\title{Magnetic resonance frequency of two-sublattice ferrimagnet \\ with magnetic compensation temperature}

\author{Kouki Mikuni\hyperlink{tokyotech}{\textsuperscript{1}}}
\author{Toshiki Hiraoka\hyperlink{tokyotech}{\textsuperscript{1}}}
\author{Takumi Kuramoto\hyperlink{ritsumeikan}{\textsuperscript{2}}}
\author{Yasuhiro Fujii\hyperlink{osakai}{\textsuperscript{3}}\hyperlink{ritsumeikan_sci}{\textsuperscript{,4}}}
\author{Akitoshi Koreeda\hyperlink{ritsumeikan}{\textsuperscript{2}}}
\author{Sergii Parchenko\hyperlink{XFEL}{\textsuperscript{5}}}
\author{Andrzej Stupakiewicz\hyperlink{Bialystok}{\textsuperscript{6}}}
\author{Takuya Satoh\hyperlink{tokyotech}{\textsuperscript{1}}\hyperlink{chirality}{\textsuperscript{,7}}}

\affiliation{\hypertarget{tokyotech}{\textsuperscript{1}Department of Physics, Institute of Science Tokyo, Tokyo 152-8551, Japan}}
\affiliation{\hypertarget{ritsumeikan}{\textsuperscript{2}Department of Physical Sciences, Ritsumeikan University, Kusatsu, Shiga 525-8577, Japan}}
\affiliation{\hypertarget{osaka}{\textsuperscript{3}Institute for Open and Transdisciplinary Research Initiatives, Osaka University, Suita, Osaka 565-0871, Japan}}
\affiliation{\hypertarget{ritsumeikan_sci}{\textsuperscript{4}Research Organization of Science and Technology, Ritsumeikan University, Kusatsu, Shiga 525-8577, Japan}}
\affiliation{\hypertarget{XFEL}{\textsuperscript{5}European XFEL, Holzkoppel 4, 22869 Schenefeld, Germany}}
\affiliation{\hypertarget{Bialystok}{\textsuperscript{6}Faculty of Physics, University of Bialystok, 1L Ciolkowskiego, 15-245 Bialystok, Poland}}
\affiliation{\hypertarget{chirality}{\textsuperscript{7}Quantum Research Center for Chirality, Institute for Molecular Science, Okazaki, Aichi 444-8585, Japan}}

\date{\today }% It is always \today, today,

\begin{abstract}
    Ferrimagnetic materials with a compensation temperature have recently attracted interest because of their unique combination of ferromagnetic and antiferromagnetic properties. 
    However, their magnetization dynamics near the compensation temperature are complex and cannot be fully explained by conventional ferromagnetic resonance (FMR) or exchange resonance modes.
    Therefore, practical models are necessary to capture these dynamics accurately.
    In this study, we derived the analytical solutions for the magnetic resonance frequencies of compensated ferrimagnets over all temperature ranges, considering both the in-plane and out-of-plane orientations of the magnetization.
    Our solutions successfully reproduce the experimental data obtained from time-resolved magneto-optical Faraday rotation and Brillouin light scattering measurements for the in-plane and out-of-plane cases, respectively.
    This reproduction is achieved by incorporating the exchange stiffness and temperature dependence of the magnetic anisotropy into the free energy density.
    Additionally, at temperatures sufficiently far from the compensation temperature, our analytical solutions converge with the conventional FMR and exchange resonance models.
\end{abstract}

\maketitle

%\tableofcontents

\section{I\lowercase{ntroduction}}
\begin{figure*}[!t]
  \centering
  \includegraphics[width=\linewidth]{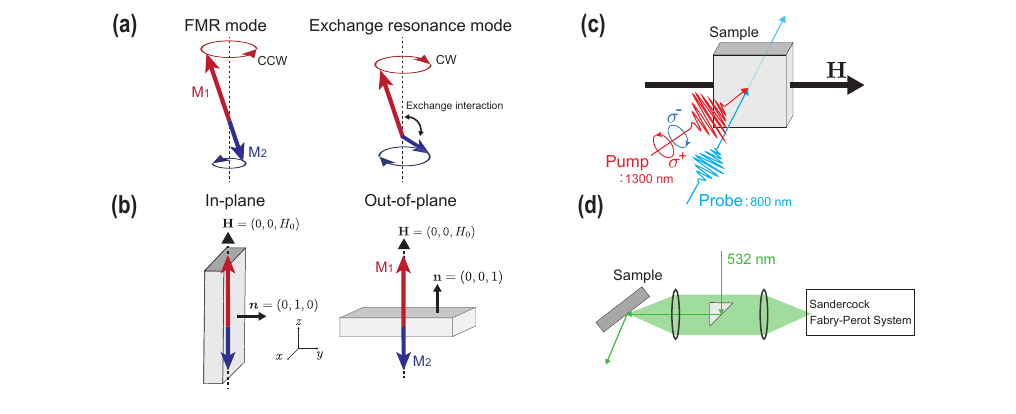}
  \caption{
    (a) Two types of magnetic resonance modes in two-sublattice ferrimagnets: FMR mode with counterclockwise (CCW) rotation (left) and exchange resonance mode with clockwise (CW) rotation (right).
    The red and blue arrows represent $\mathbf{M}_1$ and $\mathbf{M}_2$, respectively.
    (b) Coordinate system used in the calculations for in-plane (left) and out-of-plane (right) orientations. 
    (c) Experimental setup for magneto-optical pump-probe measurements with an in-plane magnetic field.
    (d) Experimental setup for Brillouin light scattering measurements.
  }
  \label{model:setup}
\end{figure*}
Ferrimagnets have been extensively studied in spintronics because of their unique properties \cite{wolf_ferrimagnetism_1961,ivanov_ultrafast_2019,finley_spintronics_2020,kim_ferrimagnetic_2022,zhang_ferrimagnets_2023}.
They are characterized by multiple magnetic sublattices with antiparallel orientations.
Owing to the unequal magnitudes of these sublattice magnetizations, ferrimagnets exhibit a net magnetization.
This unique configuration gives rise to two magnetic resonance modes in the two-sublattice ferrimagnet: 
the ferromagnetic resonance (FMR) mode in the GHz range and exchange resonance mode, which arises from the exchange interactions, in the (sub-)THz range \cite{kaplan_exchange_1953,geschwind_exchange_1959}, as shown in Fig.\ \ref{model:setup}(a).
Ferrimagnets combine the high-speed operation characteristic of antiferromagnets with the ease of control of ferromagnets owing to the net magnetization. This combination enables the use of ferromagnetic control techniques, making ferrimagnets essential for both applications and fundamental research.
Experimentally, the magnetic dynamics of ferrimagnets has been observed using THz and infrared  magnetic resonance \cite{sievers_far_1963,yamamoto_far-infrared_1974,kang_far-infrared_2012,dutta_experimental_2024}, Brillouin light scattering (BLS) \cite{borovik-romanov_brillouin-mandelstam_1982,matsumoto_optical_2018,kim_distinct_2020,haltz_quantitative_2022} and time-resolved magneto-optical Faraday rotation  \cite{stanciu_ultrafast_2006,reid_optical_2010,parchenko_non-thermal_2016,stupakiewicz_ultrafast_2021,hiraoka_sublattice-selective_2024}.

Recently, ferrimagnets with magnetization compensation temperatures ($T_\mathrm{M}$), where the net magnetization becomes zero owing to the different temperature dependences of the sublattice magnetizations, have gained attention. 
Investigations near $T_\mathrm{M}$ include spin-orbit torque-induced magnetization switching \cite{finley_spin-orbit-torque_2016,mishra_anomalous_2017}, ultrafast magnetization reversal by femtosecond laser pulses \cite{stanciu_subpicosecond_2007,radu_transient_2011,graves_nanoscale_2013,mangin_engineered_2014}, and dynamics in the noncollinear phase \cite{becker_ultrafast_2017,davydova_htextensuremath-t_2019,blank_thz-scale_2021,krichevsky_unconventional_2023,parchenko_transient_2023}.

Ferrimagnetic materials with $T_\mathrm{M}$ exhibit significantly different magnetization dynamics near $T_\mathrm{M}$ than materials without a compensation temperature.
Kaplan and Kittel\cite{kaplan_exchange_1953}, Geschwind and Walker\cite{geschwind_exchange_1959}, and Kittel\cite{kittel_theory_1959} derived the expressions for the FMR and exchange resonance modes.
Using their models, Stanciu $\mathit{et\ al.}$ \cite{stanciu_ultrafast_2006} and Binder $\mathit{et\ al.}$\cite{binder_magnetization_2006} analyzed the changes in the experimentally obtained magnetic resonance frequency and Gilbert damping in compensated GdCo.
However, their model is valid only at temperatures sufficiently far from $T_\mathrm{M}$, as their approximations do not hold otherwise. 
Although Wangsness\cite{wangsness_sublattice_1953} presented a general expression valid for all temperature ranges, the expression is too complex for practical use.

To overcome these limitations, practical models have been proposed for analyzing the experimental results. These models are effective even at temperatures close to $T_\mathrm{M}$.
One such approach derives the effective magnetic field from the free energy density and solves the coupled Landau--Lifshitz--Gilbert (LLG) equations.
Kamra $\mathit{et\ al.}$\cite{kamra_gilbert_2018} derived an analytical solution for an applied perpendicular magnetic field, whereas Haltz $\mathit{et\ al.}$\cite{haltz_quantitative_2022} performed numerical calculations for in-plane magnetic fields by incorporating various parameters.
Another approach uses Euler--Lagrange equations to describe the magnetization dynamics in the noncollinear magnetic phase \cite{zvezdin_dynamics_1979,davydova_ultrafast_2019,blank_thz-scale_2021,krichevsky_unconventional_2023}.
Moreover, models using unit N$\acute{\mathrm{e}}$el vectors in an LLG-like equation have also been proposed \cite{okuno_temperature_2019,funada_enhancement_2020}.

However, several problems remain unaddressed. One is the lack of essential parameters needed to analyze certain experiments, such as demagnetizing fields and exchange stiffness. Another limitation lies in the assumption that the parameters, in particular the magnetic anisotropy, are identical for each sublattice.

In this study, we derived analytical solutions for the magnetic resonance frequencies of two-sublattice ferrimagnets with uniaxial anisotropy over all temperature ranges for both in-plane and out-of-plane magnetic orientations.
The obtained solutions can reproduce the experimental results from magneto-optical pump-probe and BLS measurements with appropriate parameters by incorporating the temperature dependence of the magnetic anisotropy and the exchange stiffness.
We adopted the free energy density model to derive these solutions, thus allowing for comparison with conventional formulas. This approach also facilitates discussions involving different sublattice parameters. 
The obtained solutions converge to the conventional FMR and exchange resonance mode equations under the assumption that the effect of the exchange interaction is sufficiently large at temperatures far from $T_\mathrm{M}$, validating the applicability of these approximations.

\section{P\lowercase{henomenological} M\lowercase{odel}}

In this section, we derive the analytical solutions of the magnetic resonance frequencies in a two-sublattice ferrimagnet from the LLG equations, which are written as
\begin{align}
  \frac{d\mathbf{M}_i}{dt} = -\gamma_i\mathbf{M}_i\times\mathbf{H}_{i}+\frac{\alpha_i}{M_i}\mathbf{M}_i\times\frac{d\mathbf{M}_i}{dt}, \label{Model:LLG_eq}
\end{align}
where $\mathbf{M}_i$ ($i=1,2$) denotes the sublattice magnetization ($M_1>M_2$); $\mathbf{H}_{i}$ denotes the effective field experienced by $\mathbf{M}_i$; $\gamma_i$ denotes the gyromagnetic ratio; $\alpha_i$ denotes the damping constant.

$\mathbf{H}_{i}$ is characterized by the free energy density $\Phi$, which is given by
\small
\begin{align}
  \Phi = &-\mathbf{M}_1\cdot\mathbf{H}_{\mathrm{0}}-\mathbf{M}_2\cdot\mathbf{H}_{\mathrm{0}}+\lambda\mathbf{M}_1\cdot\mathbf{M}_2 \notag\\
  &-K_1\frac{(\mathbf{M}_1\cdot\mathbf{n})^2}{M_1^2}
  -K_2\frac{(\mathbf{M}_2\cdot\mathbf{n})^2}{M_2^2}+2\pi(\mathbf{M}_1\cdot\mathbf{n}+\mathbf{M}_2\cdot\mathbf{n})^2 \notag
  \\
  &+\frac{A_1}{2M_1^2}(|\nabla\mathbf{M}_1|)^2+\frac{A_2}{2M_2^2}(|\nabla\mathbf{M}_2|)^2. \label{Model:FreeEnergy}
\end{align}
\normalsize
Here, $\mathbf{H}_{\mathrm{0}}$ is the external magnetic field, $\lambda\ \lparen >$0$\rparen$ is the molecular field coefficient, $\mathbf{n}$ is a unit vector perpendicular to the sample plane, $K_i$ is the uniaxial magnetic anisotropy energy density, and $A_i$ is the exchange stiffness constant \cite{kuzmin_exchange_2020,haltz_quantitative_2022}.
The effective fields are obtained by differentiating the free energy density with respect to the  sublattice magnetization, as follows:
%\vspace{-2pt}
\begin{align}
  \mathbf{H}_{i} = -\frac{\partial\Phi}{\partial\mathbf{M}_i}. \label{Model:EffectiveField}
\end{align}
%\vspace{-2pt}
We consider the undamped case, i.e., $\alpha=0$, for simplicity.
To obtain the analytical solutions of Eq.\ \eqref{Model:LLG_eq}, we employ a linear approximation assuming that the magnetization performs a small oscillation around an external magnetic field $\mathbf{H}_{\mathrm{0}}$.
When an external field is applied along the $z$-direction $\mathbf{H}_{\mathrm{0}} = (0, 0, H_{\mathrm{0}})$, the sublattice magnetization $\mathbf{M}_i$ is expressed as follows:
\vspace{-2pt}
\begin{align}
   \mathbf{M}_1 = \mqty(m_{1x}e^{i(\omega t - kz)} \\ m_{1y}e^{i(\omega t - kz)} \\ M_1) \ ,\  \mathbf{M}_2 = \mqty(m_{2x}e^{i(\omega t - kz)} \\ m_{2y}e^{i(\omega t - kz)} \\ -M_2). \label{Model:Linear_approximation}
\end{align}
It should be noted that the solution differs depending on whether the magnetization precesses perpendicularly to the sample plane or within the plane.
These cases can be represented by the unit normal vector given by $\vb{n} = (0,1,0)$ for an in-plane case and $\vb{n} = (0,0,1)$ for an out-of-plane case.
The directions of these vectors are illustrated in Fig.\ \ref{model:setup}(b).

A secular equation is derived by substituting Eqs.\ \eqref{Model:FreeEnergy}--\eqref{Model:Linear_approximation} into Eq.\ \eqref{Model:LLG_eq}.
This procedure yields the following quartic equation, which is satisfied by the resonance frequencies  (for details, refer to the Supplemental Material 1).
The solution for in-plane orientation can be obtained as
\begin{widetext}
  \begin{align}
    \omega^{\mathrm{in}} = \frac{1}{\sqrt{2}}\Biggr\lbrack&\Omega^{\mathrm{in}}_1\Omega^{'\mathrm{in}}_1+\Omega^{\mathrm{in}}_2\Omega^{'\mathrm{in}}_2-2\gamma_1\gamma_2\lambda^2 M_1M_2 \notag
    \\
  &\pm \sqrt{(\Omega^{\mathrm{in}}_1\Omega^{'\mathrm{in}}_1-\Omega^{\mathrm{in}}_2\Omega^{'\mathrm{in}}_2)^2 - 4\gamma_1\gamma_2\lambda^2 M_1M_2(\Omega^{\mathrm{in}}_1 + \Omega^{'\mathrm{in}}_2)(\Omega^{\mathrm{in}}_2 + \Omega^{'\mathrm{in}}_1)}\Biggr\rbrack^{\frac{1}{2}}.
    \label{Model:inplane}
  \end{align}
Here, we introduce the following notations:
\begin{align*}
  \Omega^{\mathrm{in}}_1 &\coloneqq \gamma_1\qty[H_{\mathrm{0}} + \lambda M_2 - \frac{2K_1}{M_1} + 4\pi M_1 + \frac{A_1}{M_1}k^2], \ \ \ \ 
  \Omega^{'\mathrm{in}}_1 \coloneqq \gamma_1\qty[H_{\mathrm{0}} + \lambda M_2 + \frac{A_1}{M_1}k^2],\\
  \Omega^{\mathrm{in}}_2 &\coloneqq \gamma_2\qty[H_{\mathrm{0}} - \lambda M_1 +\frac{2K_2}{M_2} -4\pi M_2 - \frac{A_2}{M_2}k^2], \ \ \ \ 
  \Omega^{'\mathrm{in}}_2 \coloneqq \gamma_2\qty[H_{\mathrm{0}} - \lambda M_1 - \frac{A_2}{M_2}k^2].
\end{align*}

The calculation for the out-of-plane orientation can be performed in a similar manner (see Supplemental Material 1).
The solution, which is consistent with the result of Ref. \cite{kamra_gilbert_2018}, is 
\begin{align}
  \omega^{\mathrm{out}} = \frac{1}{2}\Biggr\lbrack\pm(\Omega^{\mathrm{out}}_1+\Omega^{\mathrm{out}}_2) + \sqrt{(\Omega^{\mathrm{out}}_1-\Omega^{\mathrm{out}}_2)^2 - 4\gamma_1\gamma_2\lambda^2M_1M_2}\Biggr\rbrack, \label{Model:outofplane}
\end{align}
where
\begin{align*}
  \Omega^{\mathrm{out}}_1 \coloneqq \gamma_1\qty[H_{\mathrm{0}} - 4\pi (M_{1}-M_{2}) + \lambda M_{2} + \frac{2K_1}{M_1} + \frac{A_1}{M_1}k^2],\\
  \Omega^{\mathrm{out}}_2 \coloneqq \gamma_2\qty[H_{\mathrm{0}} - 4\pi(M_{1}-M_{2}) - \lambda M_{1} - \frac{2K_2}{M_2} - \frac{A_2}{M_2}k^2].
\end{align*}
\end{widetext}
Regarding the $\pm$ signs on the right side of Eqs. \eqref{Model:inplane} and \eqref{Model:outofplane}, the minus sign corresponds to the low-frequency (LF) mode whereas the plus sign corresponds to the high-frequency (HF) mode.
As described in Section 4.3, at temperatures far from $T_\mathrm{M}$, the LF and HF modes correspond to the conventional FMR and exchange resonance modes, respectively.
The sublattice magnetizations rotate counterclockwise (CCW) and clockwise (CW) in the above modes, respectively [Fig. \ref{model:setup}(a)].
Equations \eqref{Model:inplane} and \eqref{Model:outofplane} correspond to a simplified version of the solution presented in Ref.\cite{wangsness_sublattice_1953}.

\section{S\lowercase{ample} \lowercase{and} M\lowercase{ethods}}
\begin{figure*}[!t]
  \centering
  \includegraphics[width=\linewidth]{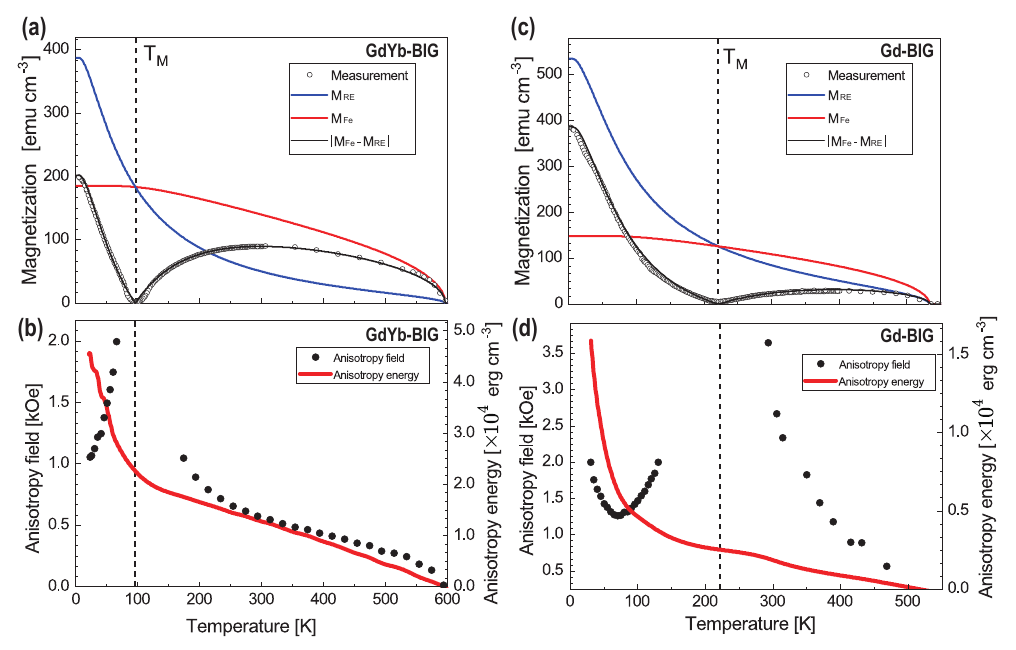}
  \caption{Temperature dependence of the sublattice magnetization for (a) GdYb-BIG (c) Gd-BIG. The open black circles represent the net magnetization measured by the superconducting quantum interference device (SQUID), and solid lines represent the calculation results of the molecular field theory.
  Temperature dependence of the uniaxial magnetic anisotropy energy for (b) GdYb-BIG (d) Gd-BIG.
  The red solid lines represent the interpolated values obtained by converting the experimentally determined anisotropy field using $H^{\mathrm{FMR}}_{\mathrm{u}} = \frac{4K}{M_1-M_2}$.}
  \label{sample:magnetization_anisotropy}
\end{figure*}

\subsection{Sample properties}
We focus on two types of bithmuth-doped rare earth (RE) iron garnets with different $T_\mathrm{M}$.
The first is \ce{Gd_{3/2}Yb_{1/2}BiFe5O12} (GdYb-BIG) with Curie temperature $T_c=573$ K and $T_{\mathrm{M}}=96$ K \cite{satoh_directional_2012,parchenko_magnetization_2014}.
Samples with thicknesses of 140 $\mathrm {\mu m}$ and 390 $\mathrm{\mu m}$ were utilized for the pump-probe and BLS, respectively.
The second is \ce{Gd_{1.8}Tb_{0.2}BiAl_{0.4}Fe_{4.6}O_{12}} (Gd-BIG) with $T_c=532$ K and $T_{\mathrm{M}}=221$ K \cite{parchenko_non-thermal_2016}.
The thicknesses of this sample were 200 $\mathrm{\mu m}$ and 475 $\mathrm{\mu m}$ for the pump-probe and BLS, respectively.
All samples are single crystals grown by the liquid-phase epitaxy method and with a (111) plane orientation.

Our RE iron garnets consist of tetrahedral and octahedral Fe sublattices and an RE magnetic sublattice.
Two Fe sublattices can be considered as one because the coupling between them is orders of magnitude stronger than the coupling between those of Fe and RE.
Additionally, in GdYb-BIG, the magnetization of the Yb ions is small when compared to that of the Gd ions and is negligible above 50 K.
From this perspective, the system can be treated as a two-sublattice ferrimagnet, consisting of RE magnetization ($\mathbf{M}_\mathrm{RE}$) and Fe magnetization ($\mathbf{M}_\mathrm{Fe}$).
We assumed that $K_1=K_2=K, A_1=A_2=A$ and the temperature-dependent parameters are $\mathbf{M}_\mathrm{RE}$, $\mathbf{M}_\mathrm{Fe}$, and $K$. 
The sublattice magnetizations were calculated from the molecular field theory [Figs.\ \ref{sample:magnetization_anisotropy}(a) and \ref{sample:magnetization_anisotropy}(c)].
The molecular field parameters were determined to reproduce the saturation magnetization measured using a vibrating-sample magnetometer and SQUID.
In the calculations, $\mathbf{M}_i$ is defined as $\mathbf{M}_1=\mathbf{M}_\mathrm{Fe}$ and $\mathbf{M}_2=\mathbf{M}_\mathrm{RE}$ for $T>T_\mathrm{M}$, and $\mathbf{M}_1=\mathbf{M}_\mathrm{RE}$ and $\mathbf{M}_2=\mathbf{M}_\mathrm{Fe}$ for $T<T_\mathrm{M}$.
$K$ is obtained from the coercivity assumed as an effective anisotropy field $H^{\mathrm{FMR}}_{\mathrm{u}}$.
$H^{\mathrm{FMR}}_{\mathrm{u}}$ diverges at $T_{\mathrm{M}}$ owing to a decrease in the net magnetization \cite{stanciu_subpicosecond_2007,p_temperature_2023}. 
To eliminate this divergence, $K$ is obtained by transforming the equation $H^{\mathrm{FMR}}_{\mathrm{u}} =\frac{4K}{M_1-M_2}$ and using interpolation [Figs.\ \ref{sample:magnetization_anisotropy}(b) and \ref{sample:magnetization_anisotropy}(d)].
Consequently, $K$ decreases monotonically as the temperature increases and is continuous even near $T_{\mathrm{M}}$.
This tendency is consistent with that reported in YIG \cite{rodrigue_resonance_1960}.
The molecular field coefficient $\lambda$ and exchange stiffness constants $A$ are the fitting parameters determined from Eqs.\ \eqref{Model:inplane} and \eqref{Model:outofplane}.
In both samples, the gyromagnetic ratios of the RE and Fe ions, which are the sublattices, are considered equal ($\gamma_{\mathrm{RE}}=\gamma_{\mathrm{Fe}}=\gamma$) above 50 K, and thus $T_{\mathrm{M}}$ and $T_{\mathrm{A}}$ coincide.

\subsection{Experimental methods}
\begin{figure*}
  \centering
  \includegraphics[width=\linewidth]{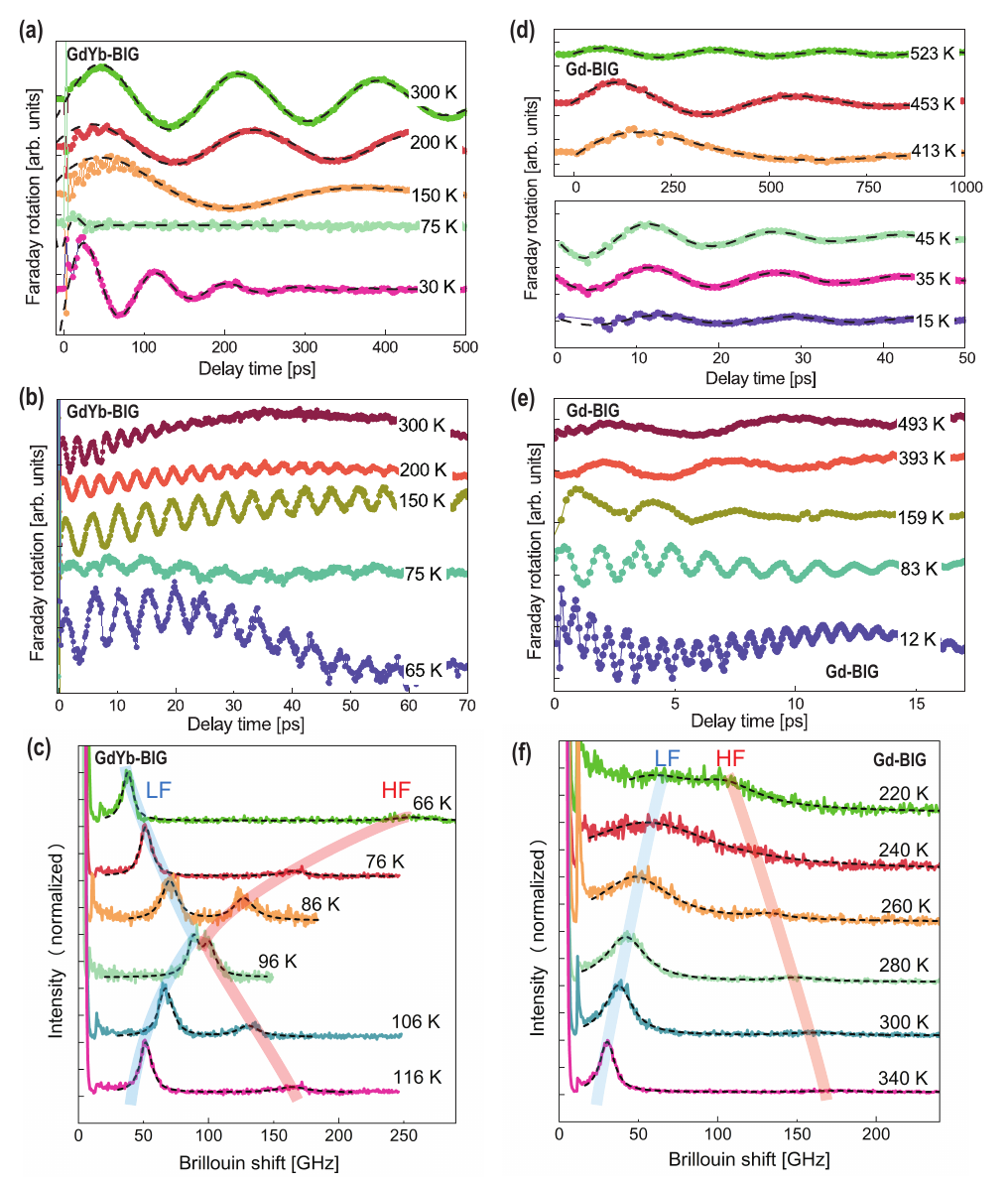}
  \caption{Time-resolved Faraday rotations at different temperatures: magnetization dynamics with (a) LF mode and (b) HF mode for GdYb-BIG, (d) LF mode and (e) HF mode for Gd-BIG. BLS results are shown for (c) GdYb-BIG and (f) Gd-BIG. The black dashed lines were fitted using Eq.\ \eqref{setup:damped ocsillation} for the pump-probe measurements and Voigt fitting for the BLS data. The BLS spectra have been normalized by the peak intensity.
  }
  \label{setup:experiment_result}
\end{figure*}
\begin{figure*}
  \centering
  \includegraphics[width=\linewidth]{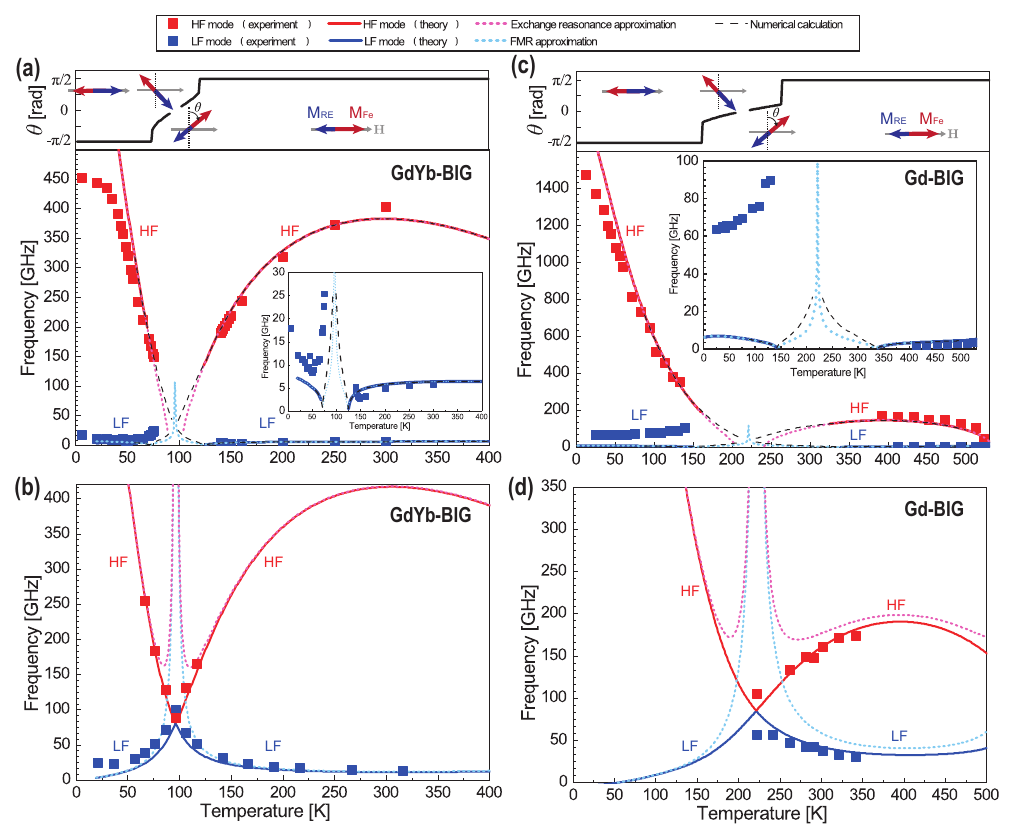}
  \caption{Temperature dependence of magnetic resonance frequencies.
   GdYb-BIG: (a) in-plane orientation, (b) out-of-plane orientation; Gd-BIG: (c) in-plane orientation, (d) out-of-plane orientation.
   The red and solid lines represent the analytical solutions of the HF mode with CW rotation and LF mode with CCW rotation, respectively [Eqs.\ \eqref{Model:inplane} and \eqref{Model:outofplane}].
   The blue and red dashed lines correspond to the FMR mode approximation and exchange resonance approximation, respectively.
   The black dashed lines in the insets represent the results of the numerical calculation. 
   In (a) and (c), the results of the numerical calculation for the magnetization at equilibrium are shown at the top of the figure.
   }
  \label{result:frequency_comparison}
\end{figure*}
For the in-plane orientation, the resonance frequency  was obtained using the magneto-optical pump-probe technique.
A circularly polarized pump pulse with a wavelength of 1300 nm and repetition rate of 500 Hz excites the magnetization dynamics via the inverse Faraday effect \cite{,kimel_ultrafast_2005,parchenko_wide_2013,parchenko_non-thermal_2016}.
A linearly polarized probe pulse with a wavelength of 800 nm and repetition rate of 1 kHz detects the Faraday rotation [Fig. \ref{model:setup}(c)].
The spot diameter of the pump pulse is approximately 280 $\mathrm{\mu m}$ with fluence of 10 $\mathrm{mJ/cm^2}$.
The externally applied in-plane magnetic field $H_{||}$ was 2 kOe for GdYb-BIG.
For Gd-BIG, the measurement of the HF mode was conducted at $H_{||}$ of 1.2 kOe  whereas that of the LF mode was conducted at $H_{||}$ of 2 kOe for  $T < T_\mathrm{M}$ and $H_{||} $ of 1.6 kOe for $T >T_\mathrm{M}$.

For the out-of-plane orientation of the magnetization, the BLS spectra were measured with a Sandercock-type tandem Fabry--Perot system using a 532 nm laser with a 100 $\mathrm{\mu m}$ spot diameter and 20 mW power.
Despite the fact that no external magnetic field was applied, the sample exhibited an out-of-plane orientation as a consequence of the influence of perpendicular uniaxial magnetic anisotropy.

\section{R\lowercase{esults} \lowercase{and} D\lowercase{iscussion}}
\begin{table}[t]
  \caption{\label{result:parameter_GdYb-BIG}%
  Material parameters used in the calculation.
  $\gamma$ and $A$ are the gyromagnetic ratio and exchange stiffness constants of each sublattice, respectively.
  $\lambda$ is the molecular field coefficient.
  The parameters are common for both the in-plane and out-of-plane orientations.
  }
  \begin{ruledtabular}
  \begin{tabular}{lccc}
  \textrm{}&
  \textrm{$\gamma$ {\scriptsize [rad/(s$\cdot$Oe)]}} &
  \textrm{$\lambda$ {\scriptsize [G/(emu/$\mathrm{cm}^3$)]}} &
  \textrm{$A$ {\scriptsize [erg/cm]}} \\ 
  \noalign{\vskip 0.7mm} % Adjust the vertical space above the line
  \hline
  \noalign{\vskip 0.7mm} % Adjust the vertical space below the line
  GdYb-BIG & $1.76 \times 10^7$  & 1550 & 5 \\
  Gd-BIG  & $1.76 \times 10^7$ & 1700 & 5 \\
  \end{tabular}
  \end{ruledtabular}
\end{table}
\subsection{In-plane orientation of magnetization}
First, we investigated the temperature dependence of the magnetic resonance frequency for the in-plane orientation.
Figures \ref{setup:experiment_result}(a) and \ref{setup:experiment_result}(b) show the time-resolved magneto-optical Faraday rotation of the magnetization dynamics with the LF and HF modes, respectively, across various temperatures in GdYb-BIG.
%Figure\ \ref{setup:experiment_result}(b) shows results in the HF mode for GdYb-BIG in the shorter time regime.
Similarly, Figs. \ref{setup:experiment_result}(d) and \ref{setup:experiment_result}(e) illustrate the LF and HF modes for Gd-BIG, respectively.
To extract the resonance frequency, the measured precession was fitted with the following damped harmonic function:
\begin{align}
  \theta_\mathrm{F}(t) = B\sin(\omega t + \phi)\exp(-\alpha\omega t) + C,
  \label{setup:damped ocsillation}
\end{align}
where $\theta_\mathrm{F}(t)$ is the Faraday rotation angle, $B$ is the amplitude, $C$ is the offset, and $\phi$ is the initial phase.

Figures \ref{result:frequency_comparison}(a) and \ref{result:frequency_comparison}(c) show the temperature dependence of the resonance frequencies extracted from the fitting using Eq. \eqref{setup:damped ocsillation} for GdYb-BIG and Gd-BIG, respectively, along with the results of the analytical solution Eq.\ \eqref{Model:inplane} shown as solid lines.
The calculated results show good agreement with the experimental data.
The parameter values used in the calculations are summarized in Table \ref{result:parameter_GdYb-BIG}.
The fitting parameters $\lambda$ and $A_i$ are comparable with those reported in Refs. \cite{Gurevich_Melkov_1996,siu_magnons_2001,liu_brillouin_1987,matsumoto_optical_2018}.
Here, the wavenumber associated with the backward volume magnetostatic wave is sufficiently small to not influence the frequencies, that is, $k=0$.

However, magnetization precession was not observed at 70--145 K for GdYb-BIG and 140--390 K for Gd-BIG even in the presence of an in-plane magnetic field.
Near $T_{\mathrm{M}}$, the influence of the magnetic anisotropy term becomes more dominant than that of the Zeeman term owing to the decrease in the net magnetization.
Consequently, the magnetization precession axis tilts perpendicular to the plane, making Eq.\ \eqref{Model:Linear_approximation} inapplicable.
This tilt reduces the excitation efficiency, as the wavevector of the pump and magnetization vector become increasingly parallel.
Additionally, the out-of-plane component of the magnetization displacement in the precession decreases.
As a result, magnetization precession is not observed near $T_{\mathrm{M}}$ even in the presence of an in-plane magnetic field.
This tilt can be confirmed by calculating the equilibrium angle $\theta$ of the magnetization, determined by the minimum of the free energy density (See Supplemental Material 2).
The top portions of Figs. \ref{result:frequency_comparison}(a) and \ref{result:frequency_comparison}(c) display the calculated equilibrium state, confirming that the magnetization is not in-plane within 74--119 K and 168--273 K for GdYb-BIG and Gd-BIG, respectively. 
To obtain accurate results within this temperature range, Eq.\ \eqref{Model:LLG_eq} is solved numerically using the magnetization equilibrium state as the initial state. 

As shown in the inset of Fig.\ \ref{result:frequency_comparison}(a), the frequency of the LF mode for GdYb-BIG has dips at 50 K and 145 K, which are below and above $T_\mathrm{M}$, respectively.
These dips are reproduced in the numerical calculation, implying the noncollinear orientation of the magnetization with respect to the external magnetic field.

Below the $T_\mathrm{M}$ for Gd-BIG, the frequency obtained in the experiment was larger than the theoretical value [Fig. \ref{result:frequency_comparison}(c)].
To investigate this discrepancy, we compared the results for $H_\mathrm{0}=10$ kOe (see Supplemental Material 3).
It was observed that the frequency of the HF mode decreased slightly with increasing $H_\mathrm{0}$, whereas that of the LF mode increased.
Therefore, we infer that the behavior of the LF mode is likely related to the FMR mode.

\subsection{Out-of-plane orientation of the magnetization}
Next, we present the results for the out-of-plane orientation.
Figures \ref{setup:experiment_result}(c) and \ref{setup:experiment_result}(f) show the BLS spectra for GdYb-BIG and Gd-BIG, respectively. 
The spectral shape was obtained as Lorentzian, assuming a Gaussian-shaped instrumental function. 
Figures \ref{result:frequency_comparison}(b) and \ref{result:frequency_comparison}(d) show the temperature dependence of the frequencies of the LF and HF modes along with the plot of Eq.\ \eqref{Model:outofplane}.
The same fitting parameters as those used for the in-plane orientation of the magnetization are applied.

The LF mode shows a frequency in tens of GHz, which is higher than that observed with the pump-probe technique.
This behavior can be reproduced by considering the backscattering magnon mode (BSM) \cite{matsumoto_optical_2018} with a wavenumber of $k_{\mathrm{BSM}} = 2nk_{\mathrm{I}} = 6.61 \times 10^7$ rad/m.
$k_{\mathrm{I}}$ is the wavenumber of the incident light, and the refractive index $n$ is 2.8 at 532 nm \cite{doormann_measurement_1984}.
Because the contribution of the BSM mode to the LF mode is large enough to reach 40 GHz, an exchange stiffness term is required to explain the observed results.

\subsection{Approximation of the analytical solution}
Equations\ \eqref{Model:inplane} and \eqref{Model:outofplane} can be approximated to represent the conventional FMR mode and exchange resonance mode at temperatures sufficiently far from $T_{\mathrm{M}}$.
This approximation is achieved by considering only the first- and second-order terms of the molecular field coefficient $\lambda$ (see Supplemental Material 4).
Ultimately, the resonance frequencies for the FMR approximation ($\omega^{\mathrm{FMR}}_\mathrm{in}$ and $\omega^{\mathrm{FMR}}_\mathrm{out}$ for the in-plane and out-of-plane orientations, respectively) and exchange resonance approximation ($\omega^{\mathrm{ex}}_\mathrm{in}$ and $\omega^{\mathrm{ex}}_\mathrm{out}$ for the in-plane and out-of-plane orientations, respectively) can be obtained as follows:
\begin{widetext}
%\footnotesize
\scriptsize
\hspace*{-0.5cm}
\begin{align}
  &\omega^{\mathrm{FMR}}_{\mathrm{in}} \sim\ \gamma_{\mathrm{eff}}^{\mathrm{FMR}}\sqrt{\qty(H_{\mathrm{0}}+H_{\mathrm{stiff}}^{\mathrm{FMR}})\qty(H_{\mathrm{0}} + 4\pi M_{\mathrm{s}} - H_{\mathrm{u}}^{\mathrm{FMR}} + H_{\mathrm{stiff}}^{\mathrm{FMR}})}
  \label{approximation:FMR_in},\\
  &\omega^{\mathrm{FMR}}_{\mathrm{out}} \sim\ \gamma_{\mathrm{eff}}^{\mathrm{FMR}}\qty(H_{\mathrm{0}}-4\pi M_{\mathrm{s}}+H_{\mathrm{u}}^{\mathrm{FMR}}+H_{\mathrm{stiff}}^{\mathrm{FMR}})
  \label{approximation:FMR_out},\\
  &\omega^{\mathrm{ex}}_{\mathrm{in}} \sim\ \lambda(\gamma_2M_1-\gamma_1M_2) -\gamma_{\mathrm{eff}}^{\mathrm{ex}}H_{\mathrm{0}} + \frac{1}{2(\gamma_2M_1-\gamma_1M_2)}\qty{4\pi M_1M_2(\gamma_1-\gamma_2)^2-\gamma_1^2\qty(\frac{2K_1}{M_1}+\frac{A_{1}}{M_1}k^2)M_2+\gamma_2^2\qty(\frac{2K_2}{M_2}+\frac{A_{2}}{M_2}k^2)M_1}
  \label{approximation:ex_in},\\
  &\omega^{\mathrm{ex}}_{\mathrm{out}} \sim\ \lambda(\gamma_2M_1-\gamma_1M_2) -\gamma_{\mathrm{eff}}^{\mathrm{ex}}\qty(H_{\mathrm{0}}-4\pi M_{\mathrm{s}})+ \frac{1}{\gamma_2M_1-\gamma_1M_2}\qty{\gamma_1^2\qty(\frac{2K_1}{M_1}+\frac{A_{1}}{M_1}k^2)M_2+\gamma_2^2\qty(\frac{2K_2}{M_2}+\frac{A_{2}}{M_2}k^2)M_1},
  \label{approximation:ex_out}
\end{align}
\end{widetext}
\normalsize
where
%\footnotesize
\begin{align*}
  &M_{\mathrm{s}}=|M_1-M_2|, \
  \\
  &\gamma_{\mathrm{eff}}^{\mathrm{FMR}} = M_1-M_2/(M_1/\gamma_1 - M_2/\gamma_2),
  \\
  &H_{\mathrm{u}}^{\mathrm{FMR}} = 2(K_1+K_2)/(M_1-M_2), 
  \\
  &H_{\mathrm{stiff}}^{\mathrm{FMR}} = (A_1+A_2)k^2/(M_1-M_2),
  \\
  &\gamma_{\mathrm{eff}}^{\mathrm{ex}} = (\gamma_2^2M_1-\gamma_1^2M_2)/(\gamma_2M_1-\gamma_1M_2).
\end{align*} 
The effective gyromagnetic ratio $\gamma_{\mathrm{eff}}^{\mathrm{FMR}}$ and effective anisotropy field $H_{\mathrm{u}}^{\mathrm{FMR}}$ were employed, as they have been previously proposed in Refs.\cite{stanciu_ultrafast_2006,binder_magnetization_2006,wangsness_sublattice_1953} and Refs. \cite{geschwind_exchange_1959,lax1962microwave}, respectively.
The FMR approximations $\omega^{\mathrm{FMR}}_{\mathrm{in}}$ and $\omega^{\mathrm{FMR}}_{\mathrm{out}}$ are consistent with the Kittel equation \cite{kittel_theory_1948}.
Additionally, the dominant term of both $\omega^{\mathrm{ex}}_{\mathrm{in}}$ and $\omega^{\mathrm{ex}}_{\mathrm{out}}$, $\lambda(\gamma_2M_1 - \gamma_1M_2)$, aligns with the equation formulated by Geschwind and Walker \cite{geschwind_exchange_1959}.

\vspace{0pt}
These approximations agree with the analytical solutions Eqs.\ \eqref{Model:inplane} and \eqref{Model:outofplane} far from $T_\mathrm{M}$, as shown in Fig.\ \ref{result:frequency_comparison}. However, as the temperature approaches $T_\mathrm{M}$, they diverge owing to the decrease in the net magnetization $M_{\mathrm{s}}$ in the denominator.
Therefore, we emphasize that the simpler formulas Eqs.\ \eqref{approximation:FMR_in},\ \eqref{approximation:FMR_out},\ \eqref{approximation:ex_in}, and \eqref{approximation:ex_out} can be used instead of the analytical solutions.
It should be noted that in the simplified case at $T_\mathrm{M}$, if we assume $\gamma=\gamma_1=\gamma_2$, $M=M_1=M_2$, and $K=K_1=K_2$, the expressions coincide with the antiferromagnetic resonance frequency formula \cite{keffer_theory_1952}:
\begin{align}
  \omega^{\mathrm{AFM}}_{\mathrm{out}}= \gamma\qty[\pm H_\mathrm{0} + \sqrt{H_\mathrm{u}(H_\mathrm{u} + 2\lambda M)}],
\end{align}
where $H_\mathrm{u} = \frac{2K}{M}$.

\subsection{Case of $\gamma_1 \neq \gamma_2$}
Thus far, we have focused on the garnet samples for which $\gamma_1=\gamma_2$, i.e., $T_{\mathrm{M}}$ coincides with the angular momentum compensation temperature $T_{\mathrm{A}}$.
However, our theoretical calculations [Eqs.\ \eqref{Model:inplane}, \eqref{Model:outofplane}, and \eqref{approximation:FMR_in}--\eqref{approximation:ex_out}] are also applicable to more general cases.
Because $T_{\mathrm{A}}$ is defined as the temperature at which $\frac{M_1}{\gamma_1} = \frac{M_2}{\gamma_2}$, $T_{\mathrm{A}}$ is not aligned with $T_{\mathrm{M}}$ when $\gamma_1 \neq \gamma_2$.
This discrepancy means that although the net magnetization becomes zero, the total angular momentum does not vanish.
These temperatures can be controlled by varying the relative composition of the sublattices \cite{dionne_molecular_1971,kim_distinct_2020}.
This characteristic is utilized in the fast domain wall motion \cite{kim_fast_2017,caretta_fast_2018} and skyrmion Hall effect \cite{kim_self-focusing_2017,hirata_vanishing_2019}.

Equations \eqref{Model:inplane} and \eqref{Model:outofplane} remain valid even when $\gamma_1 \neq \gamma_2$.
Regarding the approximations, $\gamma_{\mathrm{eff}}^{\mathrm{FMR}}$ vanishes at $T_{\mathrm{M}}$ and diverges at $T_{\mathrm{A}}$, causing $\omega^{\mathrm{FMR}}_\mathrm{in}$ and $\omega^{\mathrm{FMR}}_\mathrm{out}$ to exhibit the same behavior.
On the other hand, the first term of $\omega^{\mathrm{ex}}_\mathrm{in}$ and $\omega^{\mathrm{ex}}_\mathrm{out}$ becomes zero at  $T_{\mathrm{A}}$, as mentioned in Ref. \cite{stanciu_ultrafast_2006}.
However, in higher-order approximations, the frequency does not exactly vanish at $T_{\mathrm{A}}$. Moreover, as mentioned throughout this paper, the dynamics near $T_{\mathrm{A}}$ also require a detailed analysis to capture the full picture.

\section{C\lowercase{onclusion}}
This study aimed to obtain practical solutions for the resonance frequencies applicable across all temperature ranges including the compensation temperatures.
We proposed a two-sublattice ferrimagnet model based on the LLG equation, utilizing the free energy density that incorporates the Zeeman term, demagnetizing field term, exchange interaction, uniaxial magnetic anisotropy, and exchange stiffness.
Analytical solutions under linear approximation were derived for both in-plane and out-of-plane orientations.

These solutions successfully describe the experimental results from pump-probe measurement and BLS with appropriate parameters in the M || H region. 
Notably, exchange stiffness was crucial in explaining the increase in the LF mode of BLS.
At temperatures far from the compensation temperature, the solutions can be made consistent with conventional FMR and exchange resonance modes by assuming that the contribution of the exchange interaction dominates the other terms.
This model is also expected to be applicable to other ferrimagnets, as well as to the $M \nparallel H$ configuration by numerical calculations. 
It should be noted that the parameters for each sublattice might differ in such cases.

\section*{A\lowercase{cknowledgments}}
We thank Hiro Munekata and Kihiro T.\ Yamada for the valuable discussions and technical support.
This study was financially supported by JSPS KAKENHI (Grant Nos. JP19H01828, JP19H05618, JP19K21854, JP21H01018, JP21H01032, and JP22H01154), the Frontier Photonic Sciences Project of NINS, (Grant Nos. 01212002 and 01213004), OML Project of NINS (Grant No. OML012301), MEXT X-NICS (Grant No. JPJ011438), and JST CREST.
\bibliographystyle{myjabbrv4.bst}
\bibliography{reference}% Produces the bibliography via BibTeX.
\end{document}